\documentclass[12pt, a4paper]{article}
\usepackage{amsmath, exscale}
\allowdisplaybreaks
\DeclareMathOperator{\diag}{diag}

\let\theta=\vartheta
\begin{document}
\title{A note on post-Riemannian structures of spacetime}
\author{Friedrich W.~Hehl\thanks{E-mail address:
\texttt{hehl@thp.uni-koeln.de}} \and Uwe Muench\thanks{E-mail address:
\texttt{muench@thp.uni-koeln.de}}\\[1ex]
\small\it Institute for Theoretical Physics, University of Cologne\\
\small\it D-50923 K\"oln, Germany}
\date{}
\maketitle
\begin{abstract}
A four-dimensional differentiable manifold is given with an arbitrary
linear connection
\(\Gamma_\alpha{}^\beta=\Gamma_{i\alpha}{}^\beta\,dx^i\).
Megged~\cite{megged} has claimed that he can define a metric
\(G_{\alpha\beta}\) by means of a certain integral equation such that
the connection is compatible with the 
metric. We point out that Megged's implicite definition of his metric
\(G_{\alpha\beta}\) is equivalent to the assumption of a vanishing
nonmetricity. Thus his result turns out to be trivial.
\end{abstract}

In the metric-affine theory of gravitation \cite{hehl1, gronwald},
spacetime is assumed to be a four-dimensional differentiable manifold
equipped with a linear connection \(\Gamma_\alpha{}^\beta\) and,
\emph{independently}, with a metric \(g_{\alpha\beta}\). Ne'eman and
one of us proposed methods 
\cite{hehl2, neeman} how one could measure the torsion \(T^\alpha :=
D\theta^\alpha\) (here \(\theta^\alpha\) is the coframe) and the
nonmetricity \(Q_{\alpha\beta}:=-Dg_{\alpha\beta}\) of spacetime. In
a recent note, Megged \cite{megged} has claimed that the use of the
nonmetricity \(Q_{\alpha\beta}\) is misleading in some sense, since
one can define a metric, provided the connection is given, such that it
is automatically ``connection compatible'', if we use his words. We
will point out that this rests on the hidden assumption  of a
vanishing nonmetricity.  
Since Megged discards the original independent metric \(g_{\alpha\beta}\)
as a meaningful physical field, there is no way that nonmetricity could
enter his theory later. 

\section{Metric-affine spacetime}
If a metric \(g_{\alpha\beta}\) and a connection
\(\Gamma_\alpha{}^\beta\) are given, for the
conventions see \cite{hehl1}, then we can raise and lower indices by
means of \(g_{\alpha\beta}\), such as in \(\Gamma_{\alpha\beta} :=
\Gamma_\alpha{}^\gamma g_{\gamma\beta}\), for example. With the
definition of the nonmetricity \(Q_{\alpha\beta}:=-Dg_{\alpha\beta}\),
it is straightforward to compute the symmetric part of the connection
(see \cite{schouten} or \cite[eq.~(3.10.6)]{hehl1}): 
\begin{equation}
2 \Gamma_{(\alpha\beta)} = dg_{\alpha\beta} + Q_{\alpha\beta}
\;. \label{eq:nonmetr-def} 
\end{equation}
These are 40 equations, and \emph{no relation} between metric and
connection has been assumed. 

\section{Riemann-Cartan spacetime}
If we assume the nonmetricity to vanish, then we find a Riemann-Cartan
geometry:
\begin{equation}
Q_{\alpha\beta}=0 \quad \Rightarrow \quad 2\Gamma_{(\alpha\beta)} =
dg_{\alpha\beta}\;. 
\end{equation}
We can choose the coframe to be orthonormal (orthonormal gauge),
\begin{equation}
g_{\alpha\beta} \stackrel{*}{=} o_{\alpha\beta} := \diag
(+1,-1,-1,-1)\;, 
\end{equation}
then
\begin{equation}
\Gamma_{(\alpha\beta)} \stackrel{*}{=} 0 \quad \text{or} \quad
\Gamma_{\alpha\beta} \stackrel{*}{=} - \Gamma_{\beta\alpha} \;, 
\end{equation}
i.~e., we are left with 24 independent components of a Lorentz
connection. Thus the connection one-form is \(\text{SO}(1,3)\)-valued,
describing the fact, that the scalar product of two vectors is
\emph{invariant} under parallel transport in such a spacetime.

\section{Megged's ansatz}
He allows only for a connection to be the primary geometrical
quantity. Let us call his connection
\(\hat{\Gamma}_\alpha{}^\beta\). Then he defines \emph{implicitly} a metric
\(G_{\alpha\beta} = G_{\beta\alpha}\) by the relation
\begin{equation}
\hat{\Gamma}_\alpha{}^\gamma G_{\gamma\beta} +
\hat{\Gamma}_\beta{}^\gamma G_{\alpha\gamma} = d G_{\alpha\beta}\;,
\label{eq:G-def}
\end{equation}
see his equations \cite[eq.~(7)]{megged} and
\cite[eq.~(8)]{megged}. The \(G_{\alpha\beta}\), defined by
\eqref{eq:G-def}, can be taken for raising and lowering indices, such
as in \(\hat{\Gamma}_{\alpha\beta} := \hat{\Gamma}_\alpha{}^\gamma
G_{\gamma\beta}\), for example. Then \eqref{eq:G-def} can be rewritten
as
\begin{equation}
2 \hat{\Gamma}_{(\alpha\beta)} = d G_{\alpha\beta}\;. \label{eq:megged-sc}
\end{equation}
If we compare \eqref{eq:megged-sc} with \eqref{eq:nonmetr-def}, we
recognize that the ansatz \eqref{eq:G-def}, which represents
40~independent equations, is equivalent to the assumption 
\begin{equation}
\hat{Q}_{\alpha\beta} := - \hat{D} G_{\alpha\beta} = 0 \;. 
\end{equation}
Of course, this can also be seen directly from \eqref{eq:G-def}, since
\begin{equation}
dG_{\alpha\beta} - \hat{\Gamma}_\alpha{}^\gamma G_{\gamma\beta} -
\hat{\Gamma}_\beta{}^\gamma G_{\alpha\gamma}
=:\hat{D}G_{\alpha\beta}\;. 
\end{equation}
In other words, the equation \eqref{eq:G-def}, postulated by Megged,
amounts to the assumption of a vanishing nonmetricity
\(\hat{Q}_{\alpha\beta} = 0\). And then it is not surprising to fall
back to the metric-compatible Riemann-Cartan spacetime. 

\small
\paragraph{Acknowledgments:} We are grateful to Ofer Megged for
discussions and for making his paper available to us prior to
publication.

\end{document}